\documentclass[aps,prb,floatfix,twocolumn,superscriptaddress]{revtex4-2}
\usepackage{amsmath,amssymb}
\usepackage{graphicx}
\usepackage{epsfig}
\usepackage{epstopdf}
\usepackage[usenames]{color}
\usepackage{verbatim}
\usepackage{psfrag}
\usepackage{xcolor}
\usepackage{hyperref}
\usepackage{dcolumn}
\usepackage{bm}
\usepackage{stmaryrd} 
\usepackage{bbm}

\hypersetup{colorlinks=true,citecolor={blue},linkcolor={blue},urlcolor={blue}}

\begin{document}

\title{Impurity-band optical transitions in two-dimensional Dirac materials \\
under strain-induced synthetic magnetic field}



\author{M.~V.~Boev}
\affiliation{Rzhanov Institute of Semiconductor Physics, Siberian Branch of Russian Academy of Sciences, Novosibirsk 630090, Russia}

\author{I.~G.~Savenko}
\affiliation{Center for Theoretical Physics of Complex Systems, Institute for Basic Science (IBS), Daejeon 34126, Korea}

\author{V.~M.~Kovalev}
\affiliation{Rzhanov Institute of Semiconductor Physics, Siberian Branch of Russian Academy of Sciences, Novosibirsk 630090, Russia}

\date{\today}

\begin{abstract}
We develop a theory of optical transitions in Coulomb impurity-doped two-dimensional transition metal dichalcogenide monolayers and study the transitions from the spin-resolved valence band to the (Coulomb) donor and acceptor impurities under the influence of a synthetic valley-selective magnetic field produced by a mechanical strain.
It is shown that the optical properties of the system are determined by the strength of the synthetic magnetic field, which uncovers an experimental tool, which can be used to manipulate the properties of two-dimensional materials in valley magneto-optoelectronics. 
\end{abstract}

\maketitle


\section{Introduction}
Transition metal dichalcogenides (TMDs) belong to the family of novel truly two-dimensional (2D) monomolecular-layer materials, which have recently been attracting exceptionally large attention due to their unique  properties~\cite{XiaoVHE}. 
A typical example of a TMD, which is frequently discussed in the literature and used in experiments, is molybdenum disulfide (MoS$_2$) -- a 2D direct-bandgap semiconductor, possessing hexagonal crystal lattice structure and D$_{3h}$ point symmetry group. 
The Brillouin zone of MoS$_2$ contains two nonequivalent valleys $K$ and $K'$ coupled by the time-reversal symmetry, which defines specific optical and transport properties of this material and opens a possibility for applications in nanoelectronics, mesoscopic physics, and optoelectronics. 

In particular, the interband optical transitions in MoS$_2$ obey the valley-dependent optical selection rules: if the external electromagnetic (EM) field is circularly polarised, the interband transitions dominantly occur in one of the valleys~\cite{Zeng_2012, Kovalev_2018}. 
This feature of TMD monolayers, in general, defines their photo-induced transport properties, especially the valley Hall effect~\cite{Mak1489, RefOurPhotVHE, Jin893, GlazGolArEl, RefOurVAE,RefOurAMEWeiss}, and constitutes the concept of valleytronics~\cite{PhysRevLett.99.236809}. %
Moreover, electrons in TMDs experience relatively strong spin-orbit interaction (SOI) due to the occupied d-orbitals. 
This property (in addition to the presence of the gap in the spectrum) makes transport properties of TMDs different from graphene, in which the SOI is relatively weak. 
Moreover, large SOI creates a sufficient spin-resolved splitting of TMDs valence band~\cite{Silva_Guill_n_2016, RefKormanyos, PhysRevB.88.045416}.

The optical selection rules for interband transitions usually work well since the electron momentum is a good quantum number.
Indeed, even though each band consists of discrete k-states, the distance between them is very small and thus, the spectrum can always be treated as continuous.
However, the physical properties of semiconductors are often defined by the properties of defects~\cite{doi:10.1063/5.0018557, PhysRevLett.121.167402} and impurities residing in them~\cite{PhysRevLett.89.226601, PhysRevLett.109.245501, PhysRevB.45.11324, PhysRevB.97.165305, PhysRevB.51.14147}, in particular, the donor and acceptor centers, which become ionized at finite temperatures and enhance the conductivity. 
Their presence results in the emergence of discrete quantum states in the band gap in the vicinity of the conduction band or the valence band.
If the frequency of the external light is smaller than the bandgap energy, optical transitions can occur from the impurity state to a band~\cite{Li1993}.
The theoretical description of the properties of optical transitions from bands to impurity states and back and the analysis of the optical characteristics of these transitions (which depend on the valley quantum number) is an important problem in valley-optoelectronics of two-dimensional semiconductors. 
In particular, the selection rules here can be different from the interband transitions since one deals with a set of discrete states, characterized by the radial and angular momentum quantum numbers~\cite{PhysRevB.103.L161301}, whereas the translational momentum represents a bad quantum number due to the localization of electrons on impurities. 

In addition to fundamental interest and the modification of generic optical properties of semiconductors, impurities are important for other applications. 
For example, recently it was suggested that point-like atomic defects in TMD monolayers can potentially be utilized as single-photon emitters~\cite{doi:10.1063/5.0018557}. 
Special attention has been paid to artificially-created atomic \textit{vacancies}, and discussed the possibilities to implement them into photonic and optoelectronic devices. 
In particular, it was shown that an external applied strain can lead to a considerable modification of the optical properties of the system due to vacancies and substitution impurities~\cite{RefBahmani}. 
However, except the vacancies, other defects (like F-centers) may serve as photo-emitters and photo-detectors. 
And the optical properties of point-like defects (in particular, the Coulomb impurities) in TMD monolayers under the influence of strain have not been considered in literature to the best of our knowledge. 

To produce a uniform artificial magnetic field, it is requires to apply a pure shear strain of a special kind~\cite{Cazalilla2014}. 
This shear deformation produces the azimuthal-conserving perturbation and thus, the quantum states can be characterized by the angular momentum operator eigenstates. In case of any other types of deformations, it produces the non-uniform artificial magnetic and scalar fields, thus strongly complicating the theoretical analysis of the pseudo-magneto-optics of impurity states.

The theoretical analysis of these phenomena is the main goal of this paper.
We study the magneto-optical effect originating from an artificial valley-selective magnetic field emerging as a result of the strain applied to a TMD. 
The general idea to use artificial gauge fields due to strain was theoretically suggested in~\cite{RefKatsnelsonStress} and experimentally verified in~\cite{Levy544}. 
In the experiment, the artificial magnetic fields may reach $300$~Tesla. 
It allows for strain-induced driving of mobile charge carries in monomolecular semiconductors and semimetals like graphene, transition metal dichalcogenides, among other Dirac materials, and opens a new direction of research called \textit{strain engineering}. 

We demonstrate the possibility to monitor the optical properties of charge carriers localized on impurities by means of strain-induced gauge  fields, in particular, the pseudo-magnetic field. 
For that, we analyze the dependence of impurity-band optical transitions matrix elements and energy states on the pseudo-magnetic field strength at low temperatures. 
This opens a way to controllable driving of the optical transitions from the states corresponding to different impurity quantum numbers by means of the deformation.
To calculate the matrix elements and probabilities of the impurity-band transitions from spin-resolved valence band to both the donor and acceptor impurity states, we employ the model of the Coulomb impurity, and suggest a route to use these results in valley strain-engineering optoelectronics. 




\section{System Hamiltonian and the eigenvalue problem}

The Hamiltonian of a single electron in a Coulomb field reads
 \begin{gather}
 \label{EqMainHam1}
    \hat{H} =
    \left[\frac{\Delta}{2}\sigma_z + {\bf v}_\eta\hat{{\bf p}} + \frac{\alpha}{r}\hat{I}\right]\otimes\hat{I} + \frac{\lambda}{2}\eta(1-\sigma_z)\otimes \hat{s}_z,
 \end{gather}
where $\Delta$ is the TMD band gap,  $\alpha$ is the Coulomb potential strength;
${\bf v}_\eta=v(\eta\sigma_x,\sigma_y)$ is the velocity operator with $v=at/\hbar$ the Fermi velocity, $a$ the lattice parameter, and $t$ the hopping integral; $\lambda$ is the spin splitting of valence band;
$\eta=\pm 1$ is the valley index, which distinguishes the $K$ and $K'$ valleys in reciprocal space; and $\sigma_i$ are the Pauli matrices describing the triangle sablattices constituting the hexagonal lattice of transition metal dihalcogenide monolayer. 

Using the explicit form of operators in Eq.~\eqref{EqMainHam1}, we can rewrite the Hamiltonian for a given spin $s$ as  $\hat{H}_{s} = 
    \hat{H}_{s0} + \hat{H}_{i}$,
where the bare Hamiltonian written in the polar coordinates reads
 \begin{gather}\label{BareHamiltonian}
    \hat{H}_{s0} =
    \left(
        {\frac{\Delta}{2} \atop -iv e^{i\eta\phi}\left[\eta\partial_r + \frac{i}{r}\partial_\phi\right]}
        {-iv e^{-i\eta\phi}\left[\eta\partial_r - \frac{i}{r}\partial_\phi\right] \atop -\frac{\Delta}{2}+\lambda\eta s}
    \right),
 \end{gather}
while the interaction terms describing the electron-impurity interaction can be written in the universal form 
\begin{gather}\label{coulomb}
    \hat{H}_{i} =
    \left(
        {\frac{\alpha}{r} \atop 0}
        {0 \atop \frac{\alpha}{r}}
    \right).
 \end{gather}
Here and further we put $\hbar=1$ for clarity of the expressions. 

The Hamiltonian $\hat{H}_{s}$ due to its term $\hat{H}_{i}$ describes both the donor and acceptor states depending on the sign of the Coulomb strength parameter $\alpha$. 
If $\alpha<0$, it corresponds to the attracting Coulomb potential of the donor states, whereas for $\alpha>0$, it describes the acceptor energy states. 
Note, this Coulomb potential conserves the azimuthal symmetry. 
In other words, since the potential of the Coulomb impurity does not depend on the polar angle, we can write the eigenfunction in the form
 \begin{gather}
    \Psi(\phi,x) =
    \frac{1}{\sqrt{2\pi}}
    \left(
        {e^{i\eta(-1/2+m)\phi}\psi_1 (x) \atop i\eta e^{i\eta(1/2+m)\phi}\psi_2 (x)}
    \right),
 \end{gather}
where $m=\pm 1/2, \pm 3/2 ...$ is the angular momentum quantum number, and we have introduced a shorthand notation $x=(s,\eta,m;r)$. 
Here, $\psi_{1,2} (x)$ are the radial components of the spinor wave function $\Psi(\phi,x)$. 
Thus, $\psi_{1,2} (x)$ satisfy the steady-state Schr\"odinger equation,
 \begin{gather}\nonumber
    \left(
        {\frac{\Delta}{2}+\frac{\alpha}{r}-\varepsilon \atop - v\left[\partial_r + \frac{1/2 - m}{r}\right]}
        {v\left[\partial_r + \frac{1/2+ m}{r}\right] \atop -\frac{\Delta}{2}+\frac{\alpha}{r}+\eta s\lambda-\varepsilon}
    \right)
    \left(
        {\psi_1(x) \atop \psi_2(x)}
    \right)
    = 0.
 \end{gather}
The energy spectrum $\varepsilon$ defined by this equation contains both the discrete and continuous energy~\cite{LandauL4}. 
The shear strain produces an additional term in the Hamiltonian, which can be written in a similar way as an external pseudo-magnetic field, as it was theoretically predicted~\cite{RefGuinea1} and later experimentally verified~\cite{Levy544, RefLu2012}.

Such an artificial pseudo-magnetic field influences both the discrete and continuous domains of the electron energy spectrum. 
In particular, it produces a set of Landau levels with the characteristic energy distance between them proportional to the cyclotron frequency $\omega_c$ (under the condition $4v\sqrt{eBn}\ll\Delta$), in full analogy with the cyclotron effect. 
Its influence on the continuous part of the electron spectrum (in the conduction and valence bands) cannot be considered as a weak perturbation, and it has to be taken into account exactly. 
However, the region of the electron spectrum corresponding to states on donor and acceptor impurities is already discrete. 
Then, if the energy quantization produced by the Coulomb impurity potential is much stronger than the Landau quantization, the influence of strain-induced magnetic field can be disregarded (or taken into account  perturbatively).
In what follows, we will account exactly for the pseudo-magnetic field in the bands and neglect its influence on the impurity states.
To avoid confusion, we want to underline that the influence of the pseudo-magnetic field is essential for the matrix elements describing the impurity-band transitions. 
The assumptions mentioned above allow one to deal with an arbitrary relation of the distance between pseudo-Landau levels and the spin-orbit splitting in valence band.


\subsection{Electrons localised on impurities}

Here, we find the wave functions and the energy spectrum of an electron localized on the impurity using the Schr\"odinger equation. 
For these states, the energies $\varepsilon$ take discrete values in the bandgap of the TMD material, i.e. $-\Delta/2+\eta s\lambda<\varepsilon<\Delta/2$. 
To find the solution of the radial Schr\"odinger equation, we use the following ansatz for spinor components of electron wave function, 
 \begin{gather}\nonumber
    \left(
        {\psi_1(x) \atop \psi_2(x)}
    \right)
    =
    r^{-1/2+\gamma}e^{-\mu r/v}
    \left(
        {\sqrt{\Delta/2+\varepsilon-\lambda\eta s}\,f_1(x) \atop \sqrt{\Delta/2-\varepsilon}\,f_2(x)}
    \right),
 \end{gather}
where $\mu = \sqrt{(\Delta/2-\varepsilon)(\Delta/2+\varepsilon-\eta s\lambda)}$ and $\gamma = \sqrt{m^2-(\alpha/v)^2}$. 
At this point, some clarification is necessary.
From the general sense, the correctness of the solution requires the parameter $\gamma$ to be real. 
Indeed, if $\gamma$ was complex, the ground-state energy would become undefined, $\varepsilon_0\rightarrow \infty$. In other words, the electron ``falls'' on the impurity center 
but such behavior is unphysical and should be excluded from the theory.
This issue is known from the Coulomb problem in systems of Dirac particles~\cite{LandauL4}. 
It imposes a restriction on the possible values of $\alpha$: $|\alpha|/v< 1/2$. Under this condition, the system remains in the so-called \textit{subcritical regime} \cite{Novikov2007,Pottelberge2017}.

\begin{widetext}
Furthermore, the system of equations we are to solve reads
 \begin{gather}
    \left\{
    {\sqrt{\frac{\Delta/2+\varepsilon-\eta s}{\Delta/2-\varepsilon}}\left(\Delta/2-\varepsilon+\frac{\alpha}{r}\right)f_1 +
    \left[
        v\partial_r -\mu + v\frac{\gamma+m}{r}
    \right]
    f_2 = 0,
    \atop
    \left[
        v\partial_r -\mu + v\frac{\gamma-m}{r}
    \right]
    f_1 +
    \sqrt{\frac{\Delta/2-\varepsilon}{\Delta/2+\varepsilon-\eta s}}\left(\Delta/2+\varepsilon-\frac{\alpha}{r}-\eta s\lambda\right)f_2=0.}
    \right.
 \end{gather}
For convenience, we introduce functions $g_1=(f_1+f_2)/2$ and $g_2=(f_1-f_2)/2$, and variable $\rho=2\mu r/v$ yielding 
%
%
%
 \begin{gather}
    \left[
        \rho\partial^2_\rho + (1+2\gamma-\rho)\partial_\rho - \left(\gamma+\frac{\alpha(\varepsilon-\eta s\lambda/2)}{v\mu}\right)
    \right]
    g_1 = 0,
    \\
    \left[
        \rho\partial^2_\rho + (1+2\gamma-\rho)\partial_\rho - \left(1+\gamma+\frac{\alpha(\varepsilon-\eta s\lambda/2)}{v\mu}\right)
    \right]
    g_2 = 0.
 \end{gather}
%
\end{widetext}

The solutions of these equations represent Laguerre polinomials,
 \begin{gather}
 \label{EqGenSolg12}
    g_1 =
    C_1 L^{2\gamma}_{n}(\rho),
    \\
    g_2 =
    C_2 L^{2\gamma}_{n-1}(\rho),
 \end{gather}
where
 \begin{gather}\label{eq11}
    n =
    -\left[\gamma + \frac{\alpha(\varepsilon-\eta s\lambda/2)}{v\mu}\right].
 \end{gather}
We can also rewrite~\eqref{eq11} in the form
 \begin{gather}
    \mu =
    \frac{\alpha(\eta s\lambda/2-\varepsilon)}{v(n+\gamma)}
 \end{gather}
and immediately see that the electron energy $\varepsilon$ must obey the relation $\eta s\lambda/2>\varepsilon$ for acceptor impurities while $\varepsilon>\eta s\lambda/2$ for donors. 
Then, the energy levels read
 \begin{gather}\label{EnergyImp}
    \varepsilon^{(i)}_{n,m,\eta,s} 
    =\frac{\eta s\lambda}{2}-\frac{\rm{sign}(\alpha)\Delta_{\eta s}}{2}\zeta_{n,m},
 \end{gather}
where $\Delta_{\eta s}=\Delta-\eta s\lambda$ and  $\zeta_{n,m}=\left( 1+\alpha^2/[v^2(n+\gamma)^2] \right)^{-1/2}$.

The coefficients $C_2$ and $C_1$ in Eq.~\eqref{EqGenSolg12} are, in fact, not mutually independent.  
To express $C_2$ via $C_1$, we can take a particular point $r = 0$ and obtain
 \begin{gather}
    C_2 =
    - \frac{\gamma + \frac{\alpha(\varepsilon-\eta s\lambda/2)}{v\mu}}{\frac{\alpha\Delta_{\eta s}}{2v\mu}-m} \, C_1.
 \end{gather}
In the state with $n=0$, we have $\gamma =- {\alpha(\varepsilon-\eta s\lambda/2)}/{v\mu}$ and ${|\alpha|\Delta_{\eta s}}/{2v\mu} = |m|$. 
Therefore, the finiteness of the wave function of the electron on acceptor (donor) requires $n=0,1,2,...$ when $m<0$ ($m>0$) and $n=1,2,...$ when $m>0$ ($m<0$), respectively.

\begin{widetext}
The wave function of the localized electron reads
 \begin{gather}
    \Psi =
    \frac{C_1\Delta_{\eta s}^\gamma r^{-1/2+\gamma}e^{-\mu r/v}}{\sqrt{2\pi}v^\gamma}
    \left(
        {\left[\frac{\Delta_{\eta s}(1-{\rm sign}(\alpha)\zeta_{nm})}{2v}\right]^{1/2}e^{i\eta(-1/2+m)\phi}
        \left\{
            L_n^{2\gamma}(2\mu r/v) +
            \frac{nL_{n-1}^{2\gamma}(2\mu r/v)}{{\rm sign}(\alpha)\sqrt{\alpha^2/v^2+(n+\gamma)^2}-m}
        \right\}
        \atop
        i\eta\left[\frac{\Delta_{\eta s}(1+{\rm sign}(\alpha)\zeta_{nm})}{2v}\right]^{1/2} e^{i\eta(1/2+m)\phi}
        \left\{
            L_n^{2\gamma}(2\mu r/v) -
            \frac{nL_{n-1}^{2\gamma}(2\mu r/v)}{{\rm sign}(\alpha)\sqrt{\alpha^2/v^2+(n+\gamma)^2}-m}
        \right\}}
    \right).
 \end{gather}
To find $C_1$, we use the normalization condition $\int\limits_0^{2\pi} d\phi\,\int\limits_0^\infty rdr\, \Psi^+\Psi = 1$
and the relation $\int\limits_0^\infty dx\, L^\alpha_m(x)L^\alpha_n(x)x^\alpha e^{-x} =
    \frac{\Gamma(n+\alpha+1)}{n!}\delta_{m,n}$,
yielding
 \begin{gather}
    C_1 =
    \left( \frac{|\alpha|}{\sqrt{\alpha^2+v^2(n+\gamma)^2}} \right)^{1/2+\gamma}
    \left( \frac{n!}{\Gamma(n+2\gamma+1)} \right)^{1/2}
    \left[ 1+ \frac{n^3}{(n+2\gamma)\left({\rm sign}(\alpha)\sqrt{\alpha^2/v^2+(n+\gamma)^2}-m\right)} \right]^{-1/2}.
 \end{gather}
\end{widetext}


\subsection{Electrons in conduction and valence bands in pseudo-magnetic field}
In previous subsection, we found the expressions for the eigenenergies and eigenfunctions of electrons on the impurity states. 
Here, let us consider electrons in the conduction and valence bands accounting for both strain-induced pseudo-magnetic field and the spin-orbit coupling.
We start with the Hamiltonian for the  electron with spin $s$, $\hat{H}_s^\textrm{(B)} = \hat{H}_{s0} + \hat{H}_\textrm{strain}$, where the superscript ``B'' stands for ``band''. The regular bare Hamiltonian is given by Eq.~\eqref{BareHamiltonian}, while the second term describes the strain-induced electron energy and has the following form
\begin{gather}\label{eqHstr}
    \hat{H}_\textrm{strain} =
    vbr
    \left(
        -\sin(\phi)\sigma_x + \eta \cos(\phi)\sigma_y
    \right).
\end{gather}
Here, $b=2\beta_2tC/v=eB/2$ with the tight-binding model parameter $\beta_2$,  $C$ characterizes the strength of the deformation, and $B$ is the pseudo-magnetic induction~\cite{Cazalilla2014}. 
Thus, for example, for  $B=10$~T, $\beta_2=3$ and $a=3.193$~$\AA$ we find $C\approx 0.43$~$\mu$m$^{-1}$. In other words, the stress at the edge of the flake takes the value $\sigma\approx 22~L$, where we used the shear modulus ${\bar \mu}=50.4$~N/m \cite{PhysRevB.87.035423}, and the flake size $L$ is in $\mu$m.
To find the eigenfunctions and eigenstates, we separate the polar angle and the radial variables as before. 
Thus, the wave function is presented in the spinor form,
 \begin{gather}
    \Psi^{(B)}(\phi,y) =
    \frac{1}{\sqrt{2\pi}}
    \left(
        {e^{i\eta(-1+l)\phi}\psi_1^{(B)}(y) \atop i\eta e^{i\eta l\phi}\psi_2^{(B)}(y)}
    \right),
 \end{gather}
where $l$ is an integer, and we used the short-hand notation $y=(s,\eta,m;r)$.
The functions $\psi_{1,2}^{(B)}(y)$ describe the radial electron motion and obey the system of equations
 \begin{gather}\label{eqB4}
    \left(
        {\varepsilon-\frac{\Delta}{2} \atop v\left[\partial_r - \frac{l-1}{r}-br\right]}
        {-v\left[\partial_r + \frac{l}{r}+br\right] \atop \varepsilon+\frac{\Delta}{2}-\eta s\lambda}
    \right)
    \left(
        {\psi_1^{(B)}(y) \atop \psi_2^{(B)}(y)}
    \right)
    = 0.
 \end{gather}
Expressing $\psi_{1}^{(B)}(y)$ from the first line in Eq.~\eqref{eqB4} and substituting it to the second line, we find
 \begin{gather}\label{eqpsi2}
    \left[
        \xi\partial_\xi^2 + \partial_\xi - \frac{l^2}{4\xi} - \frac{l-1}{2} + \frac{\tilde{\mu}^2}{4v^2b} - \frac{\xi}{4}
    \right]
    \psi_2^{(B)}(y) = 0,
 \end{gather}
where we changed the variable  $r\rightarrow \sqrt{\xi/b}$ and introduced 
\begin{eqnarray}
\label{EqMuTwiddle}
\tilde{\mu}^2 = ({\varepsilon-\Delta/2})({\varepsilon+\Delta/2-\eta s\lambda}).
\end{eqnarray}
Furthermore, we take into account the asymptotic behavior of electron radial wave function at long and short distances, $\psi_2\sim e^{\xi/2}$ and $\sim\xi^{|l|/2}$, respectively, and find the solution of Eq.~\eqref{eqpsi2} in the form
 \begin{gather}
     \psi_2^{(B)}(y) =  e^{-\xi/2} \xi^{|l|/2} f_2^{(B)}(y),
 \end{gather}
where the function $f_2^{(B)}(y)$ obeys the equation
 \begin{gather}\label{eqf2}
    \left[
        \xi\partial_\xi^2 +\left(1+|l|-\xi\right)\partial_\xi + \frac{\tilde{\mu}^2/2v^2b-l-|l|}{2}
    \right]
    f_2 = 0.
 \end{gather}
Eq.~\eqref{eqf2} has the solution in the form of the Laguerre polynomial, $f_2^{(B)}(y)=A_2L^{|l|}_{n_2}(\xi)$, where $A_2$ is a normalization constant, and
 \begin{gather}\label{eq8}
    n_2 =
    \frac{\tilde{\mu}^2/2v^2b-l-|l|}{2}
 \end{gather}
is an integer non-negative number. To find the electron energy spectrum in a given band, we substitute Eq.~\eqref{EqMuTwiddle} into~\eqref{eq8} and introduce the principal quantum number $n=n_2+(l+|l|)/2$, yielding the expression for the electron energy in conduction $(c)$ and valence $(v)$ bands,
 \begin{eqnarray}\label{EnergyBands}
    \varepsilon_{n,\eta,s}^{(c/v)} =
    \frac{\eta s\lambda \pm \sqrt{\Delta_{\eta s}^2+16v^2bn}}{2},
 \end{eqnarray}
where the sign $+(-)$ corresponds to the energy of the conduction (valence) band.

The expression for the $\psi_{1}^{(B)}(y)$ component can be derived in the same way as the $\psi_{2}^{(B)}(y)$ component. Repeating all the steps above, we finally find the spinor components
 \begin{eqnarray}
 \label{Apsi1sol}
    \psi_1^{(B)}(y) &=&
    A_1 e^{-\xi/2} \xi^{|l-1|/2} L^{|l-1|}_{n-1-(l-1+|l-1|)/2}(\xi),
    \\
 \label{Apsi2sol}
    \psi_2^{(B)}(y) &=&
    A_2 e^{-\xi/2} \xi^{|l|/2} L^{|l|}_{n-(l+|l|)/2}(\xi).
 \end{eqnarray}
To establish a link between the coefficients $A_1$ and $A_2$, let us substitute the solutions~\eqref{Apsi1sol} and~\eqref{Apsi2sol} in Eqs.~\eqref{eqB4} after the change of variables $r\rightarrow \sqrt{\xi/b}$ and $\partial_r\rightarrow 2\sqrt{b\xi}\partial_\xi$. 
We find
 \begin{eqnarray}
 \label{A1A2}
     &&\left(
        \varepsilon - \frac{\Delta}{2}
     \right)
     \frac{\xi^{|l-1|/2}}{2v\sqrt{b}}  L^{|l-1|}_{n_1}(\xi)A_1 
     \\ 
     \nonumber
     &&~~~=\xi^{(|l|-1)/2}
     \left(
        \frac{l-|l|}{2}L^{|l|}_{n_2}(\xi) + 
        (n_2+|l|)L^{|l|-1}_{n_2}(\xi)  
     \right)A_2,
  \end{eqnarray}
\begin{eqnarray}\nonumber
     &&\xi^{(|l-1|+1)/2}
     \left(
        \frac{|l-1|-(l-1)}{2\xi}L^{|l-1|}_{n_1}(\xi) - L^{|l-1|+1}_{n_1}(\xi)  
     \right) A_1 
     \\ 
     \label{A1A2_2}
     &&~~~=\left(
        \varepsilon + \frac{\Delta}{2} - \eta s\lambda
     \right)
     \frac{\xi^{|l|/2}}{2v\sqrt{b}}  L^{|l|}_{n_2}(\xi) A_2.
  \end{eqnarray}
%
The structure of these equations suggests that we should consider three possible cases. 

First, let $n\neq 0$ and $l>0$. 
In this case, the set of equations simplifies to
 \begin{eqnarray}
    \frac{\Delta_{\eta s}\tilde{\zeta}^{\mp}_{n\eta s}}{4v\sqrt{b}}A_1 + n A_2 = 0,
    \\
    -A_1 + \frac{\Delta_{\eta s}\tilde{\zeta}^{\pm}_{n\eta s}}{4v\sqrt{b}}A_2 = 0,
 \end{eqnarray}
where $\tilde{\zeta}^{\pm}_{n\eta s}=1\pm \sqrt{1+16v^2bn/\Delta^2_{\eta s}}$, and the top (bottom) sign stands for conduction (valence) band. 
Thus, we find
 \begin{eqnarray}
    A_1 =
    \frac{2nv\sqrt{b}}{\varepsilon_n-\Delta/2} A_2 =
    - \frac{4vn\sqrt{b}}{\Delta_{\eta s}\tilde{\zeta}^{\mp}_{n\eta s}} A_2.
 \end{eqnarray}

Second, let $n\neq 0$ and $l\leq 0$.
Then, the set \eqref{A1A2}-\eqref{A1A2_2} takes the form
 \begin{eqnarray}
    -\frac{\Delta_{\eta s}\tilde{\zeta}^{\mp}_{n\eta s}}{4v\sqrt{b}}A_1 + A_2 = 0,
    \\
    n A_1 + \frac{\Delta_{\eta s}\tilde{\zeta}^{\pm}_{n\eta s}}{4v\sqrt{b}}A_2 = 0,
 \end{eqnarray}
and therefore, the relation between the coefficients is
 \begin{eqnarray}
    A_1 =
    \frac{4v\sqrt{b}}{\Delta_{\eta s}\tilde{\zeta}^{\mp}_{n\eta s}} A_2.
 \end{eqnarray}

The third case corresponds to $n=0$ and requires special consideration. 
The solution Eq.~\eqref{Apsi1sol} makes sense here only if $A_1=0$ because at $n=0$, the lower index has non-positive value resulting in non-polynomial solutions. 
Moreover, the solution Eq.~\eqref{Apsi2sol} is not divergent at infinity only if $l\leq 0$. 
It is easy to show, that Eq.~\eqref{A1A2} is satisfied for arbitrary $A_2$ and for both the bands when $n=0$, $l\leq 0$, and $A_1=0$. 
However, Eq.~\eqref{A1A2_2} gives nonzero $A_2$ only for the valence band. 
Thus, for a given deformation, the state with $n=0$ exists only in the valence band. 

\begin{figure}[t]
\includegraphics[width=0.40\textwidth]{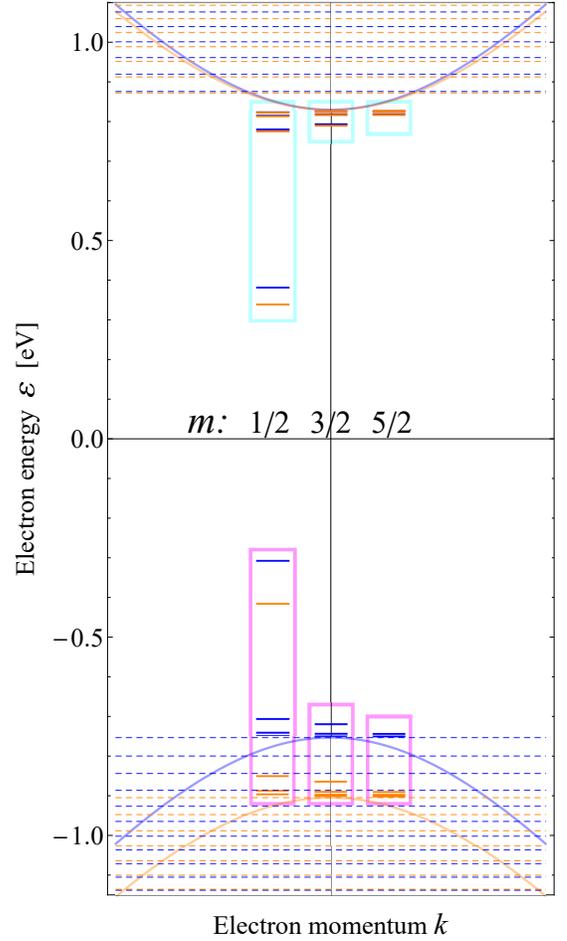}
\caption{Energy spectrum of acceptor (magenta squares) and donor (cyan squares) centers in $K$ valley;
quasi-Landau level structure of the conduction and valence bands. 
Blue dashed lines indicate the spin-up band states, while orange lines depict the spin-down states. 
We used the parameters for MoS$_2$: $\Delta =1.66$~eV, $\lambda=0.075$~eV, $B=100$~T, $t=1.1$~eV, $a=3.193$~$\AA$, $|\alpha|/v=0.45$.}
\label{Energy}
\end{figure}

Summing up, the wave function of the electron in a given band for $l>0$ and $n\neq 0$ reads
\begin{gather}
    \Psi^{(B)}(\phi,y) =\left[
        \frac{2b(n-l)!}{n!}
    \right]^{1/2}
    \left[
        1+\frac{16v^2bn}{(\Delta_{\eta s}\tilde{\zeta}_{n\eta s}^{\mp})^2}
    \right]^{-1/2}\\
    \nonumber
    \times
    \frac{b^{l/2}
    e^{-br^2/2}}{\sqrt{2\pi}}
    \left(
        {-e^{i\eta(-1+l)\phi}\frac{4vn}{\Delta_{\eta s}\tilde{\zeta}_{n\eta s}^{\mp}}r^{l-1}L^{l-1}_{n-l}(br^2) \atop i\eta e^{i\eta l\phi}r^{l}L^{l}_{n-l}(br^2)}
    \right),
 \end{gather}
and for $l\leq 0$ and $n\neq 0$ it is
\begin{gather}
    \Psi^{(B)}(\phi,y) =\left[
        \frac{2bn!}{(n+|l|)!}
    \right]^{1/2}
    \left[
        1+\frac{16v^2bn}{(\Delta_{\eta s}\tilde{\zeta}_{n\eta s}^{\mp})^2}
    \right]^{-1/2}\\
    \nonumber
    \times
    \frac{b^{l/2}
    e^{-br^2/2}}{\sqrt{2\pi}}
    \left(
        {e^{-i\eta(1+|l|)\phi}\frac{4vb}{\Delta_{\eta s}\tilde{\zeta}_{n\eta s}^{\mp}}r^{1+|l|}L^{1+|l|}_{n-1}(br^2) \atop i\eta e^{-i\eta |l|\phi}r^{|l|}L^{|l|}_{n}(br^2)}
    \right),
 \end{gather}
while for $l\leq 0$ and $n=0$ in the valence band it is given by the expression
 \begin{eqnarray}
    \Psi^{(v)}(\phi,y) =
    \frac{i\eta b^{(|l|+1)/2}}{\sqrt{\pi |l|!}}
    e^{-i\eta|l|\phi} r^{|l|}e^{-br^2/2}
    \left(
        {0 \atop 1}
    \right).
 \end{eqnarray}
%
Having solved the eigenvalue problem for the electrons in conduction and valence bands and localised on the  donor and acceptor impurity states, we can further study the optical properties of the system exposed to  light with the frequency corresponding to impurity-band transitions. %
\begin{figure}[t!]
\includegraphics[width=0.23\textwidth]{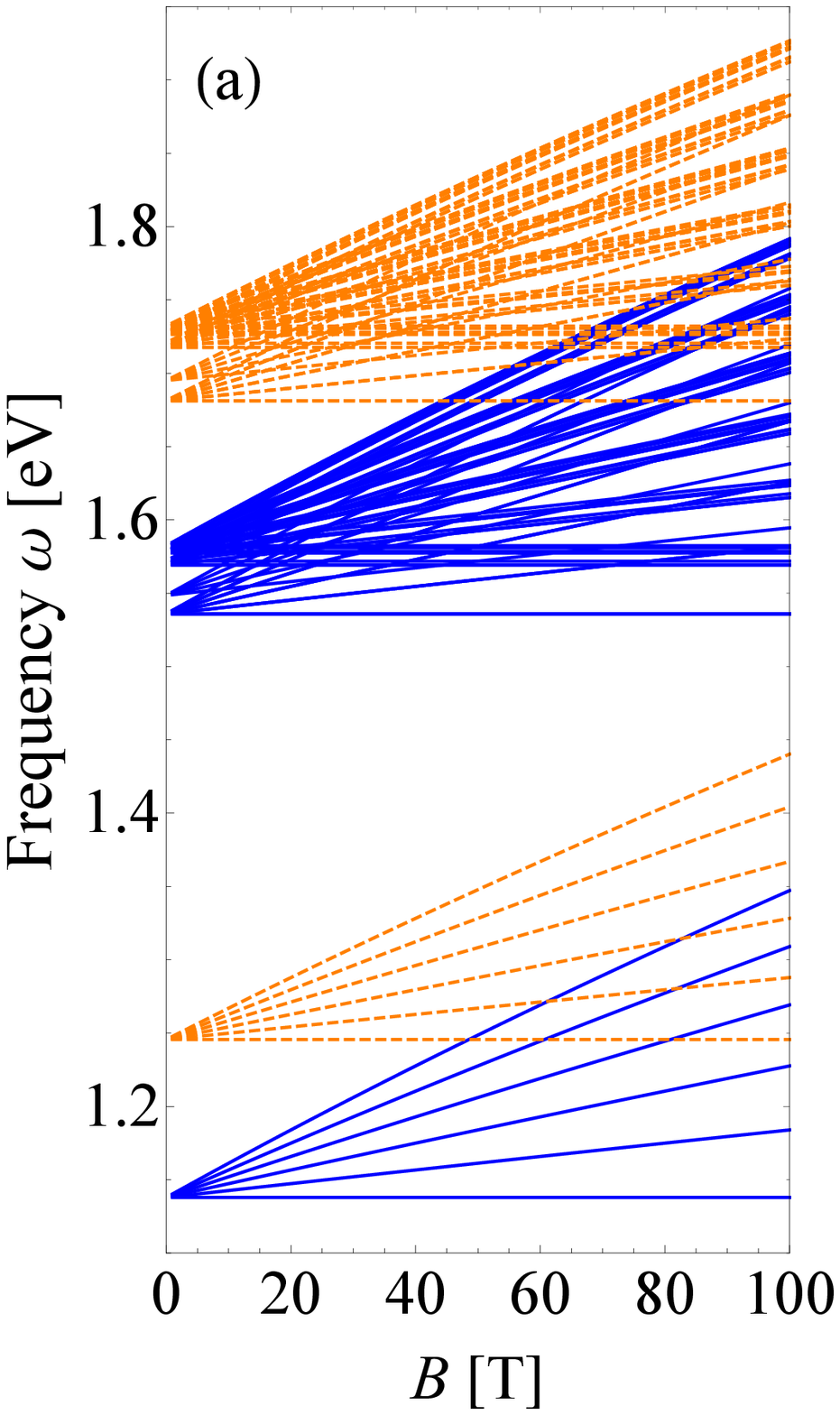}
\includegraphics[width=0.23\textwidth]{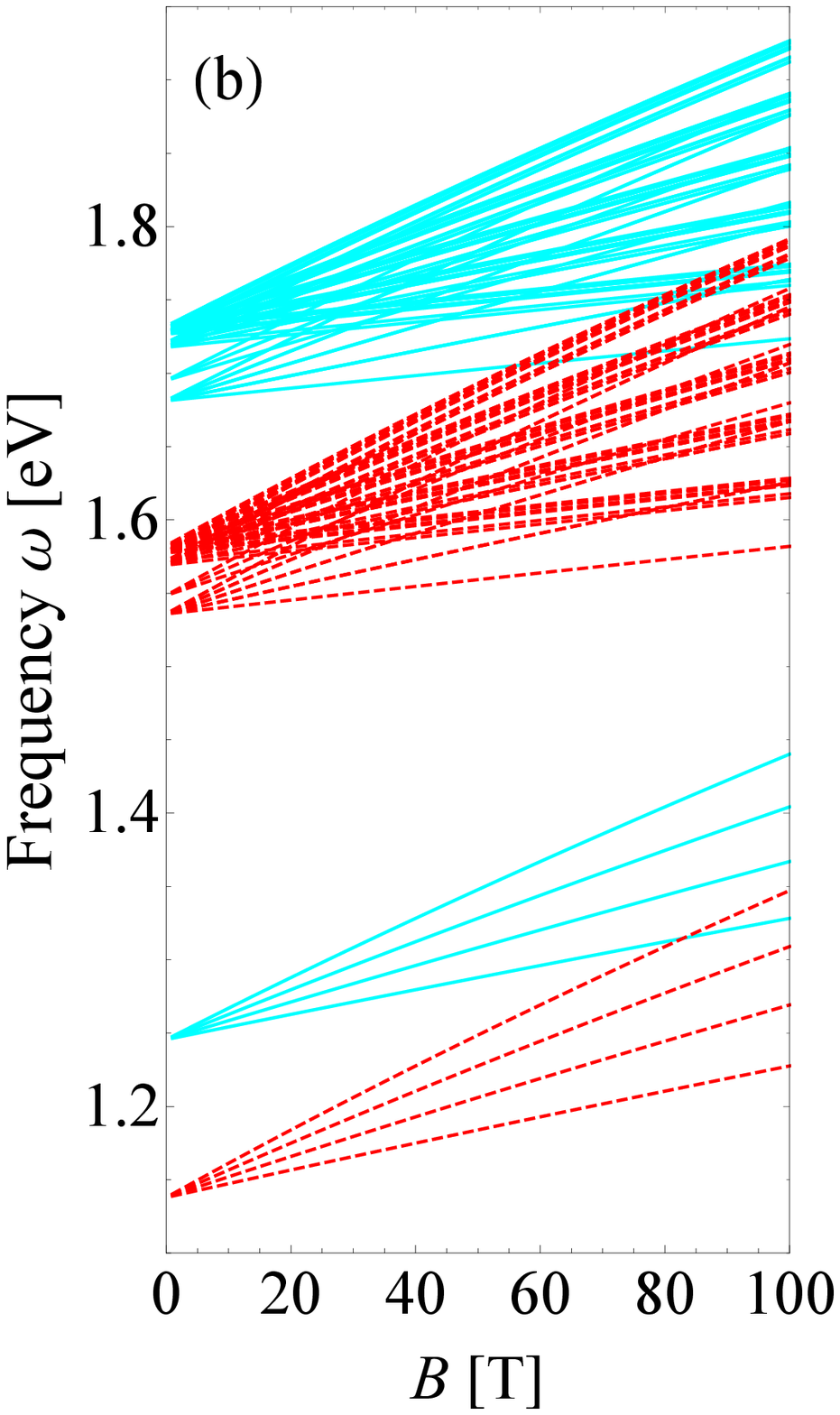}
\caption{Transitions from the valence band to donor impurity states: 
the dependence of resonant frequency $\omega$ on the pseudo-magnetic field $B$. 
We used the following quantum numbers: $n_b=[0;5]$ 
, $l=[-4;4]$, $n_i=[0,3]$, $m=[-5/2,5/2]$, $\nu=1$, $\eta= 1$ (a), $\eta= -1$ (b), $s= 1$ (solid lines), $s= -1$ (dashed lines).}
\label{Fig2}
\end{figure}

Figure~\ref{Energy} shows the impurity energy spectrum by Eq.~\eqref{EnergyImp} and the quasi-Landau levels in bands by Eq.~\eqref{EnergyBands}. 
To build the plot, we took the magnetic field strength equal to $100$ Tesla. 
This magnitude corresponds to the boundary of applicability range of our model since the Landau level splitting at this magnetic field becomes comparable with the energy splitting between the highest impurity states. 
At the same time, the distance between  the ground impurity level and other impurity high-energy states greatly exceeds the Landau-level cyclotron energy, supporting the applicability of our model, which neglects the direct influence of pseudo-magnetic field to the impurity states. 
Thus, our approach is well applicable to the lowest impurity states, which usually play the most important role in optical transitions (because the optical transitions from the higher states are usually smeared out by the temperature effects).
%
%
%
\begin{figure}[t!]
\includegraphics[width=0.23\textwidth]{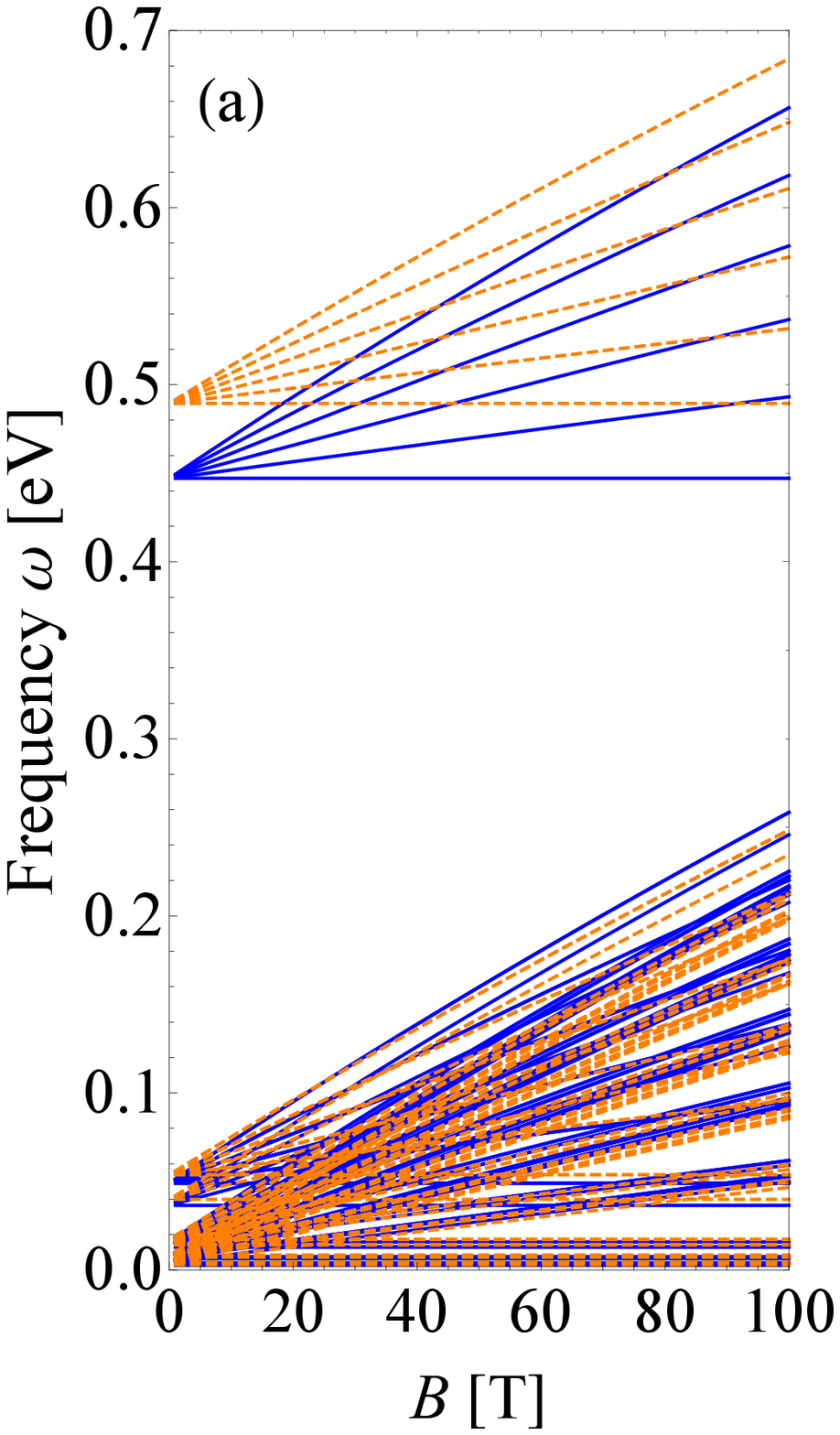}
\includegraphics[width=0.23\textwidth]{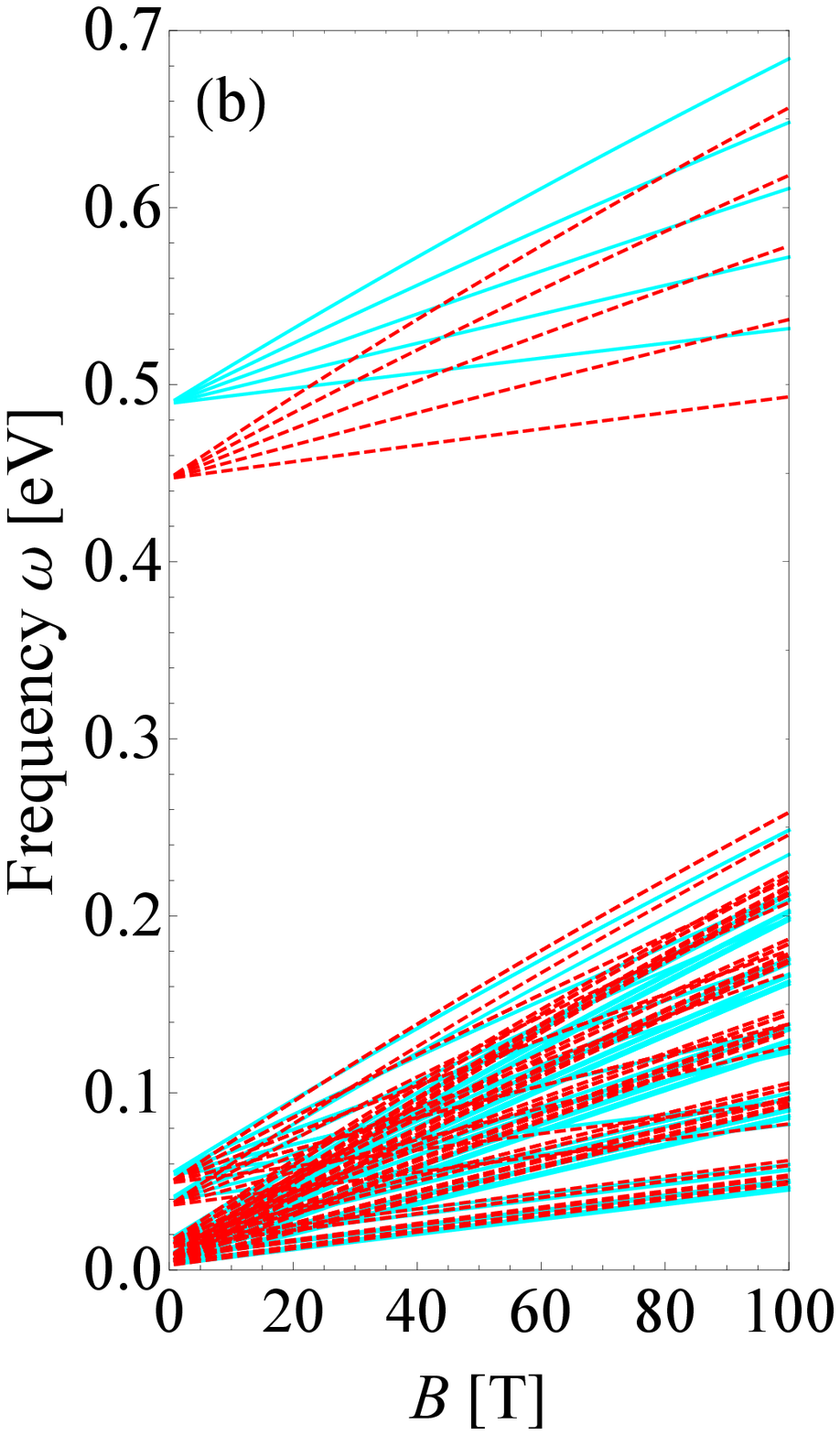}
\caption{Transitions from the valence band to acceptor impurity states: the dependence of the resonant frequency $\omega$ on the pseudo-magnetic field $B$. 
We used the following quantum numbers: $n_b=[0;5]$, $l=[-4;4]$, $n_i=[0,3]$, $m=[-5/2,5/2]$, $\nu=1$, $\eta= 1$ (a), $\eta= -1$ (b), $s= 1$ (solid lines), $s= -1$ (dashed lines).}
\label{Fig3}
\end{figure}

\section{Impurity-band optical transitions}

Let us now expose the sample to a circularly polarized EM field
 \begin{gather}
    {\bf E} =
    {E}_0 (\cos (\omega t), \nu\sin(\omega t)),
 \end{gather}
where $\nu=\pm1$ is a polarization. 
In the Hamiltonian, the interaction of the charged particle with the EM field enters as a term $\hat{{\bf j}}\cdot{\bf A}$, where
 \begin{gather}
    {\bf A} =
    - {A}_0 (\sin (\omega t), -\nu\cos(\omega t))
 \end{gather}
is the vector potential with $A_0 = - E_0/ \omega$, and we use the gauge ${\rm div}{\bf A} = 0$. 
Thus, the operator describing the interaction of electrons with the EM field reads
 \begin{gather}
    \hat{V}_{\eta,\nu}(t) =
    - e \nu{\bf v}\cdot{\bf A} =
    e\nu A_0v
    \left(
        {0 \atop -ie^{i \eta\nu\omega t}}
        {ie^{-i\eta\nu\omega t} \atop 0}
    \right).
 \end{gather}

For convenience, let us write the wave functions in a brief form
 \begin{gather}\label{eq1}
    \Psi^{(i)}_{n_i,m,\eta,s} =
    \frac{1}{\sqrt{2\pi}}
    \left(
        {e^{i\eta(-1/2+m)\phi}\psi_1^{(i)} \atop i\eta e^{i\eta(1/2+m)\phi} \psi_2^{(i)}}
    \right),
    \\
    \Psi^{(B)}_{n_b,l,\eta,s} =
    \frac{1}{\sqrt{2\pi}}
    \left(
        {e^{i\eta(-1+l)\phi}\psi_1^{(b)} \atop i\eta e^{i\eta l\phi}\psi_2^{(b)}}
    \right),
 \end{gather}
where the superscripts $B$ and $i$ denote the states in the bands and on impurities, respectively, and  
let us find the matrix element describing the transition from the valence band to an impurity state $(v)\rightarrow (i)$,
 \begin{gather}
 \label{TranMatrixElements}
    V_{(v)\rightarrow (i)}(t) =
    \left\langle
        \Psi^{(i)}_{n_i,m,\eta,s}(\phi,r;t)
        \left|
            \hat{V}_{\eta,\nu}(t)
        \right|
        \Psi^{(v)}_{n_b,\,l,\eta,s}(\phi,r;t)
    \right\rangle 
    \\
    =e^{i(\varepsilon_{n_i,m,\eta,s}^{(i)}-\varepsilon_{n_b,\eta,s}^{(v)})t}\\
    \nonumber
    ~\times
    \left\langle
        \Psi^{(i)}_{n_i,m,\eta,s}(\phi,r)
        \left|
            \hat{V}_{\eta,\nu}(t)
        \right|
        \Psi^{(v)}_{n_b,\,l,\eta,s}(\phi,r)
    \right\rangle.
 \end{gather}
Substituting here Eq.~\eqref{eq1} and integrating over the polar angle, we find $V_{(v)\rightarrow (i)}(t) =
    V_{(v)\rightarrow (i)} e^{i(\varepsilon_{n_i,m,\eta,s}^{(i)}-\varepsilon_{n_b,\eta,s}^{(v)}-\omega)t}$,
where 
\begin{eqnarray}
\label{equation}
 &&V_{(v)\rightarrow (i)}=-\nu\eta eA_0v\\
 \nonumber
 &&~~~~~~\times\left[
        \delta_{\eta,\nu}\delta_{l,m-1/2} I_1^{(v)\rightarrow (i)} +
        \delta_{\eta,-\nu}\delta_{l,m+3/2} I_2^{(v)\rightarrow (i)}
    \right]
\end{eqnarray}
and
 \begin{gather}\nonumber
 I_1^{(v)\rightarrow (i)}=\int\limits_0^\infty rdr\,\psi_1^{(i)} \psi_2^{(v)};\,\,
 I_2^{(v)\rightarrow (i)}=\int\limits_0^\infty rdr\,\psi_2^{(i)} \psi_1^{(v)}
 \end{gather}
are the overlap integrals of the  components of the electron spinor wave functions in a band and on the impurity. 

The absorption coefficient is determined by the probability of optical transitions from a given band to impurity states. 
This probability can be found from the Fermi golden rule,
 \begin{equation}
 \label{EqFermiGoldRule}
    W_{(v)\rightarrow(i)} = 
    2\pi \left|V_{(v)\rightarrow (i)}\right|^2 \delta(\varepsilon_{n_i,m,\eta,s}^{(i)}-\varepsilon_{n_b,\eta,s}^{(v)}-\omega).
 \end{equation}
In the following section, we show the results of calculations by Eq.~\eqref{EqFermiGoldRule} (and the preceding formulas).



\section{Results and discussion}
Figure~\ref{Fig2} and Fig.~\ref{Fig3} show the dependence of the resonant transition frequency [which is defined by the delta-function in Eq.~\eqref{EqFermiGoldRule}] on the artificial magnetic field strength $B$. 
As it follows from the formulas (and the figures), the resonant frequency behaves as $\sqrt{B}$, which is different from the conventional situation of materials with parabolic energy band, where the dependence of the distance between the Landau levels on the magnetic field strength is linear. 
This square-root dependence originates from  non-parabolicity of the particle spectrum typical for TMD monolayers.
It holds for both the valence band -- donor and valence band -- acceptor transitions (with a  quantitative differences only). 
We also conclude, that the EM perturbation does not produce any spin-flip processes conserving the spin quantum number under optical band-impurity transitions.  

\begin{figure}[!t]
\includegraphics[width=0.23\textwidth]{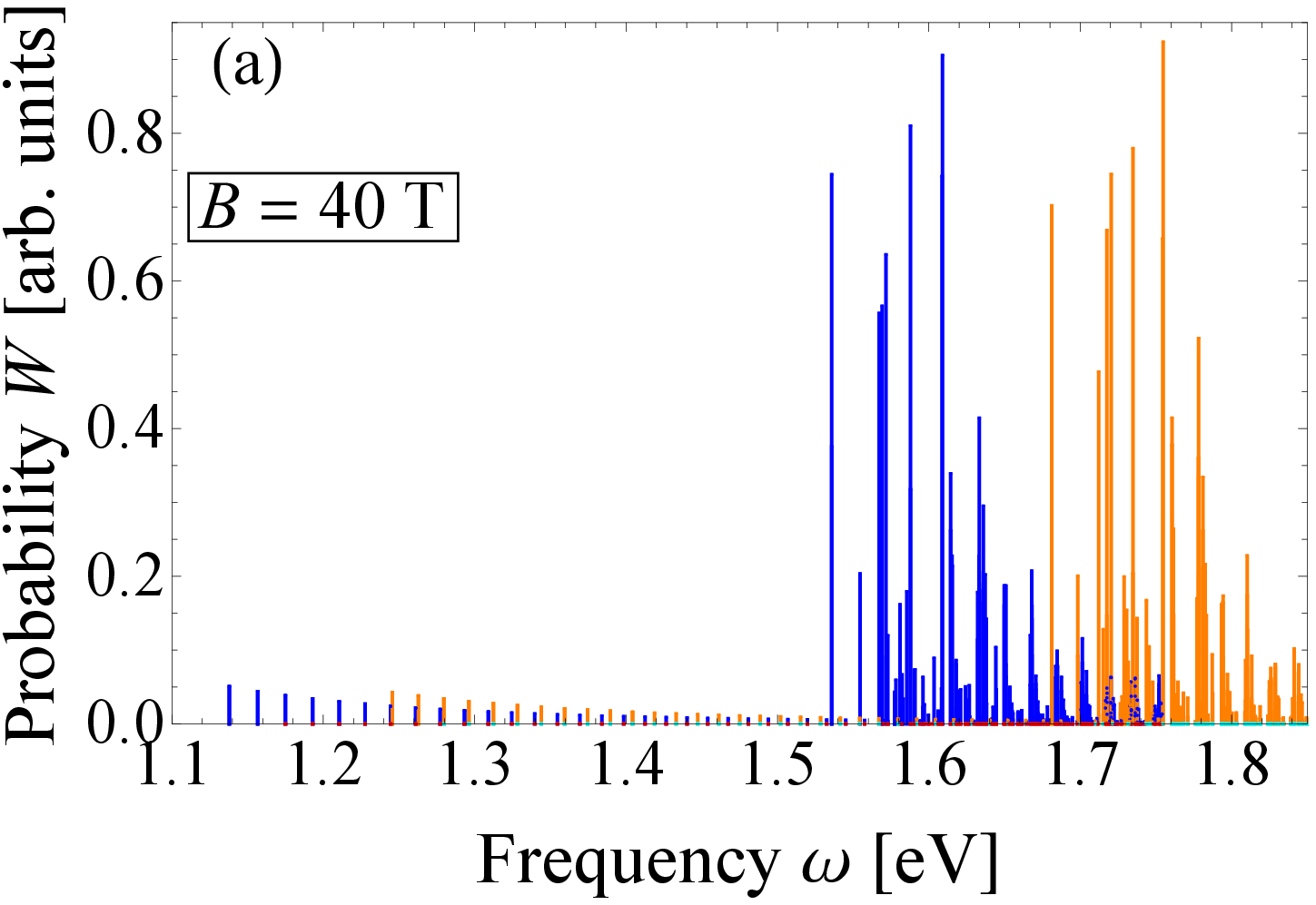}
\includegraphics[width=0.24\textwidth]{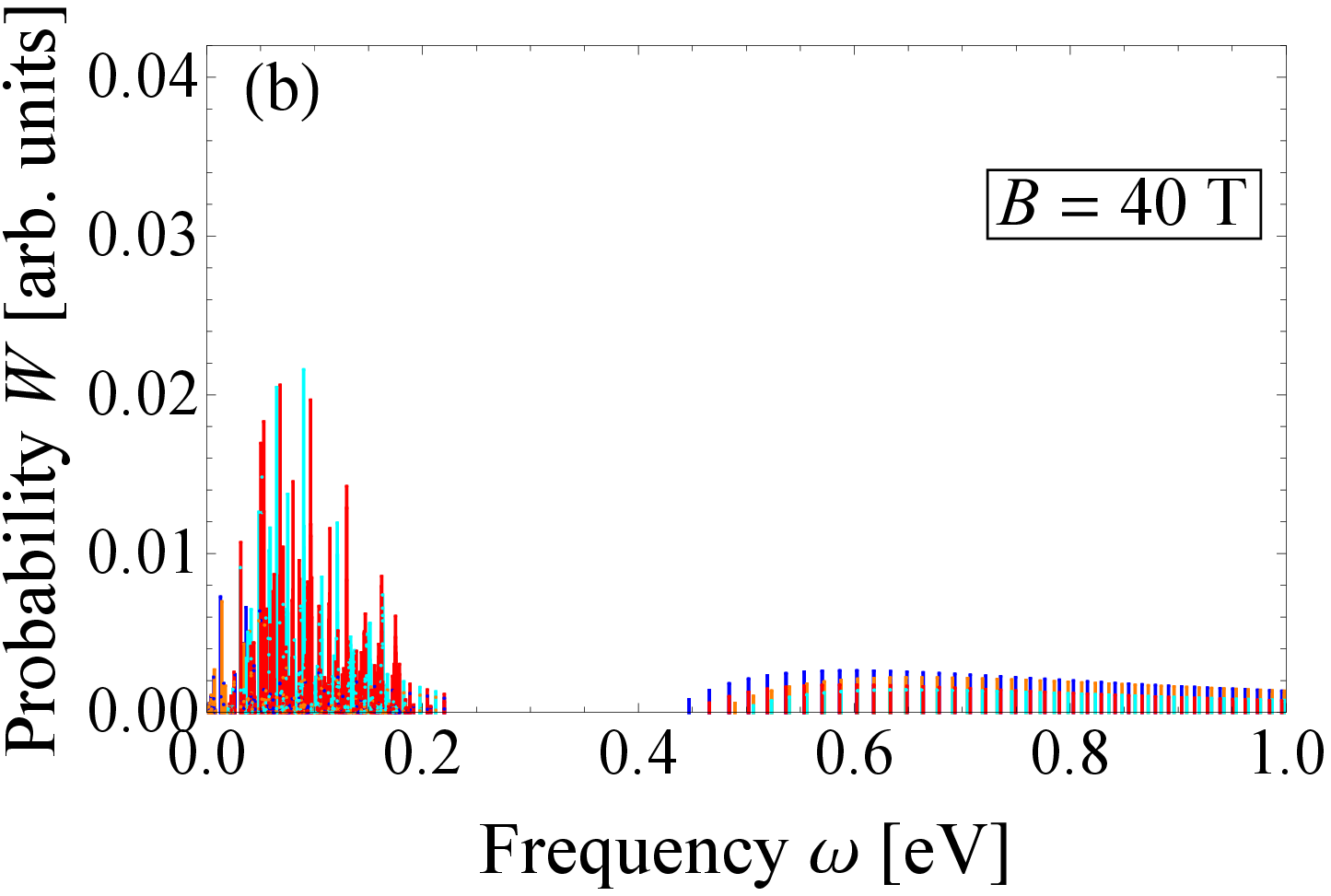}
\includegraphics[width=0.23\textwidth]{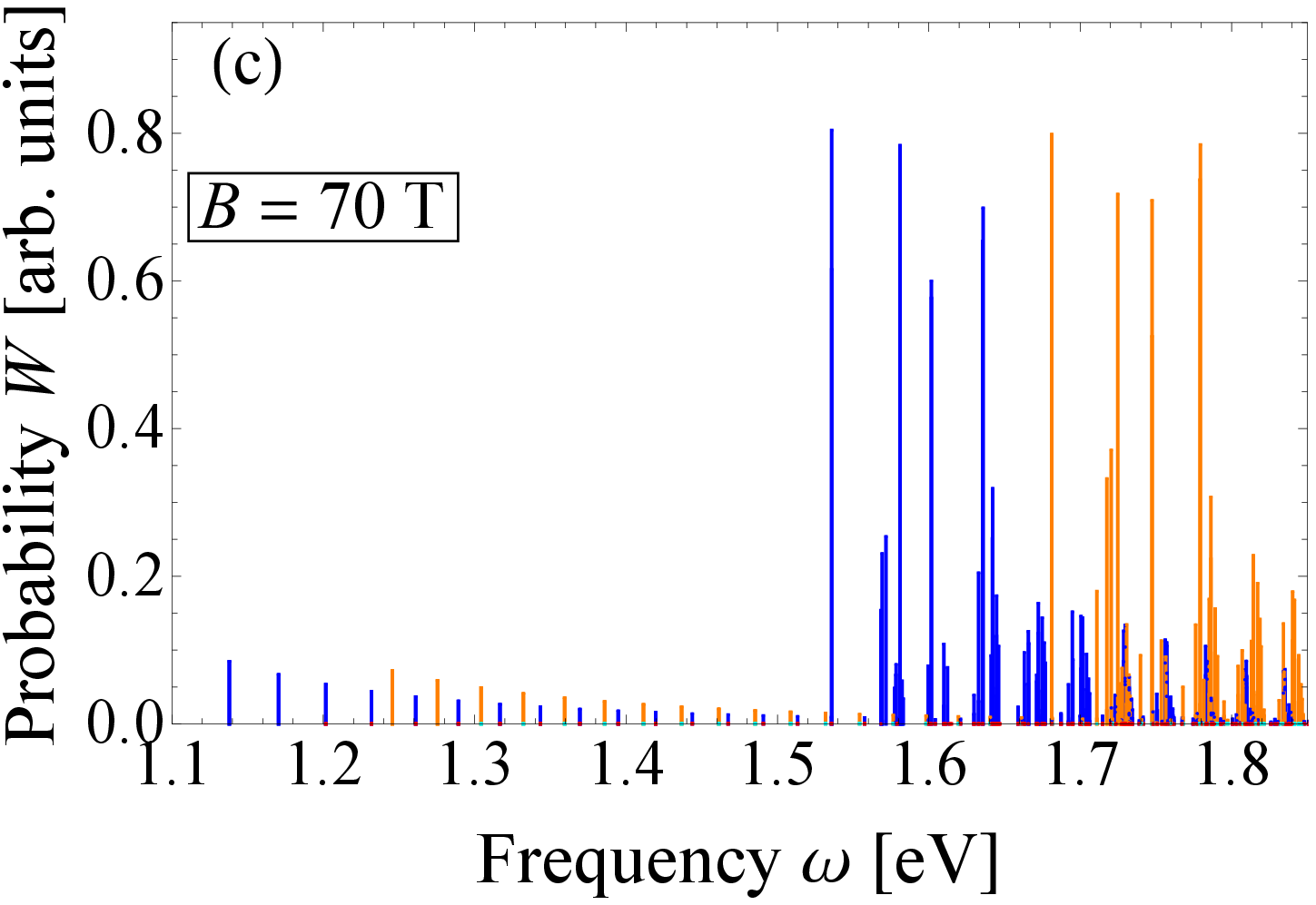}
\includegraphics[width=0.24\textwidth]{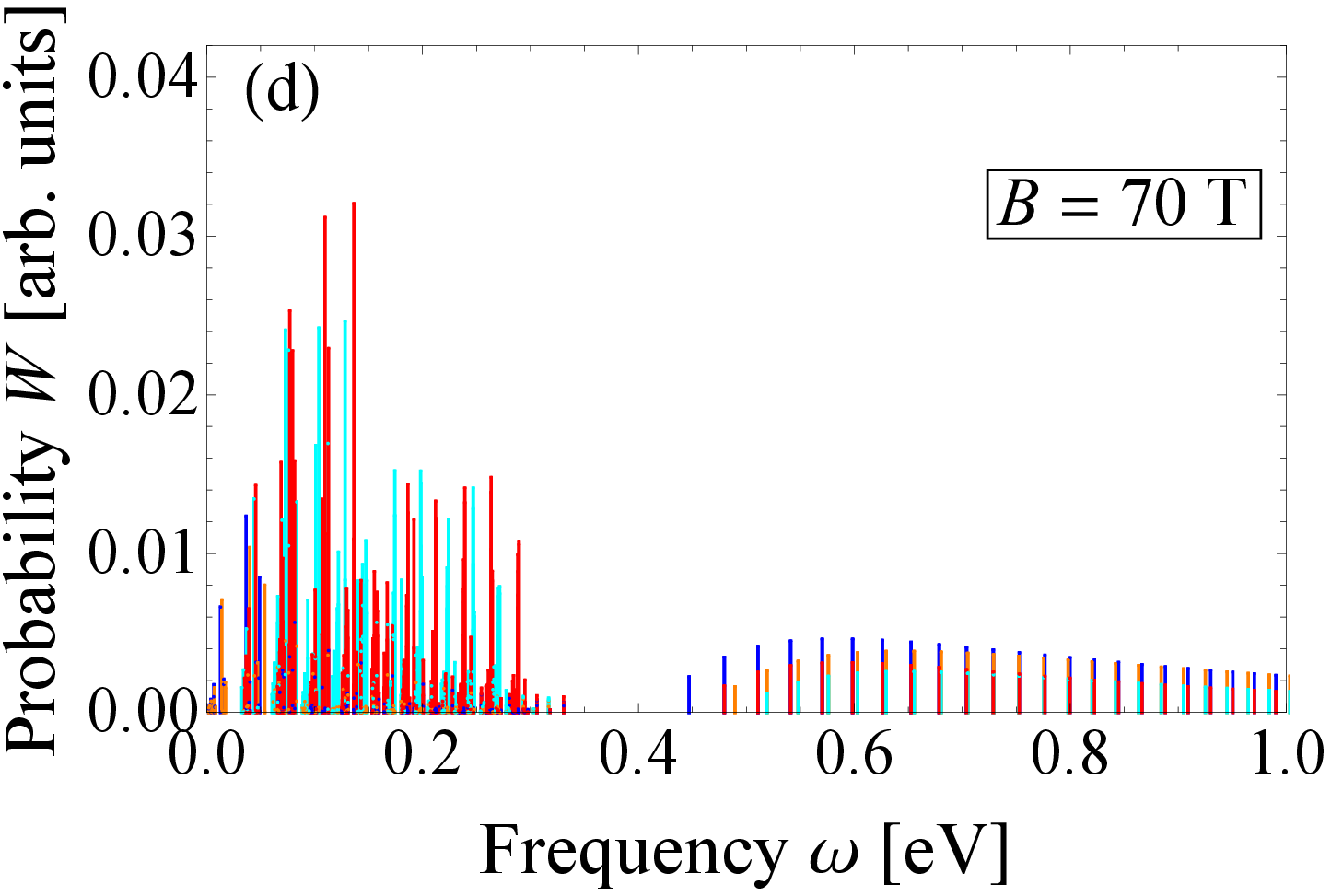}
\includegraphics[width=0.23\textwidth]{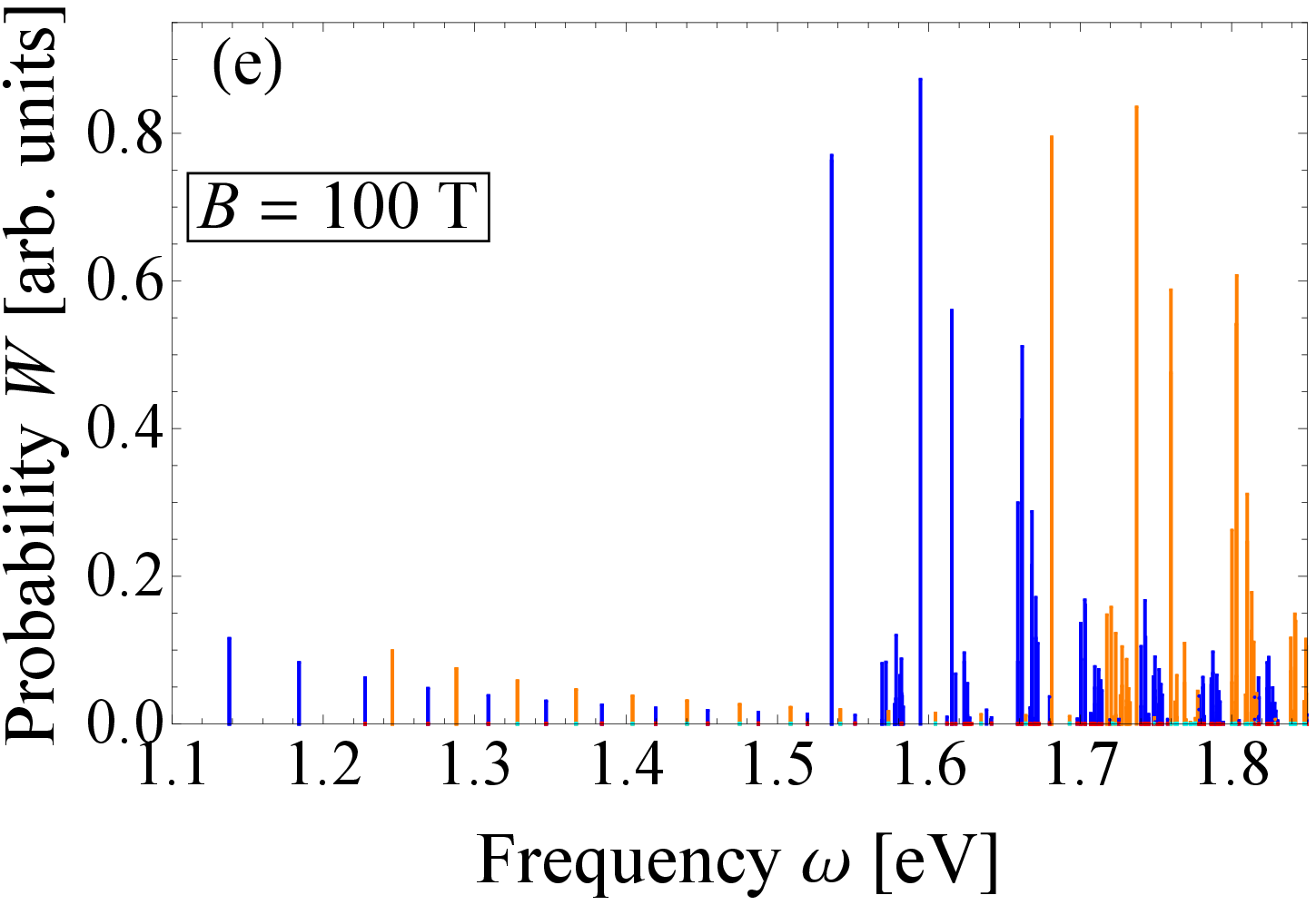}
\includegraphics[width=0.24\textwidth]{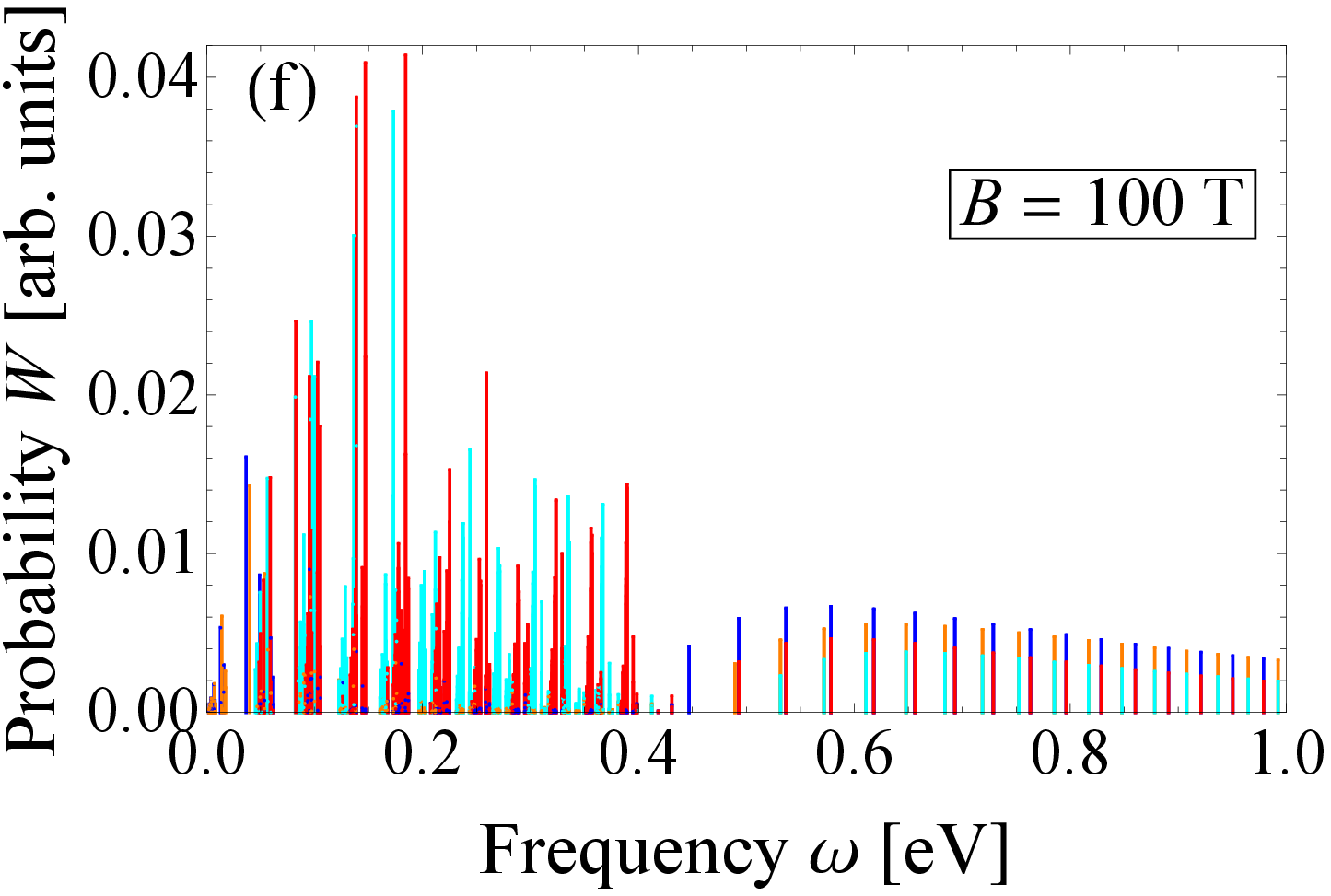}
\caption{Intensities of the optical transitions from valence band to donor (a, c, e) and acceptor (b, d, f) states for different $B$. 
The colors correspond to Fig.~\ref{Fig2} and Fig.~\ref{Fig3}.}
\label{Fig4}
\end{figure}

Let us now consider the intensities of the optical transitions. 
They are proportional to the squares of the matrix elements given in Eq.~\eqref{TranMatrixElements}. 
%
%
Figure~\ref{Fig4} demonstrates the dependence of the probabilities defined by Eq.~\eqref{EqFermiGoldRule} on the EM field frequency. 
\begin{figure}[!t]
\includegraphics[width=0.22
\textwidth]{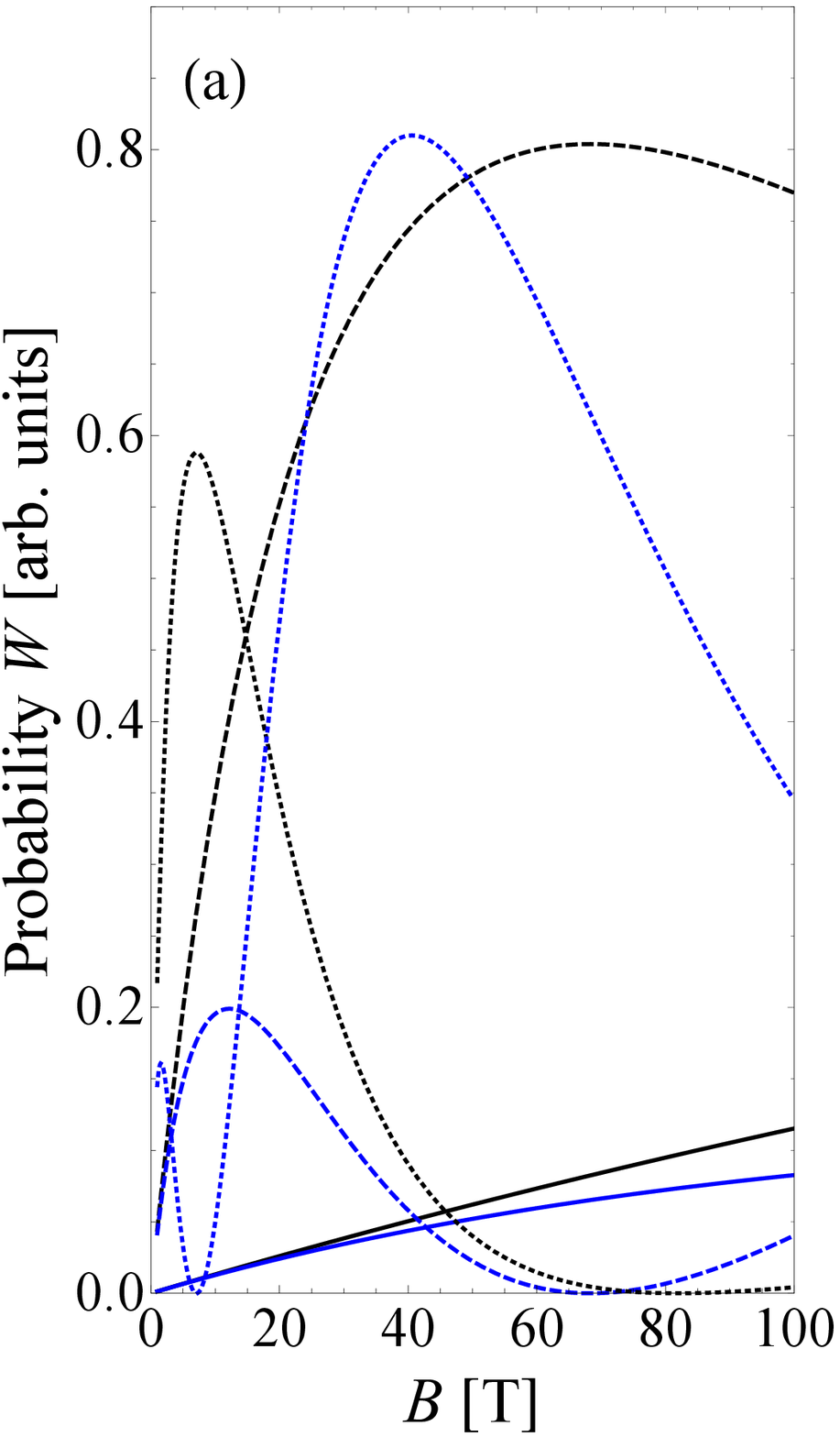}
\includegraphics[width=0.22\textwidth]{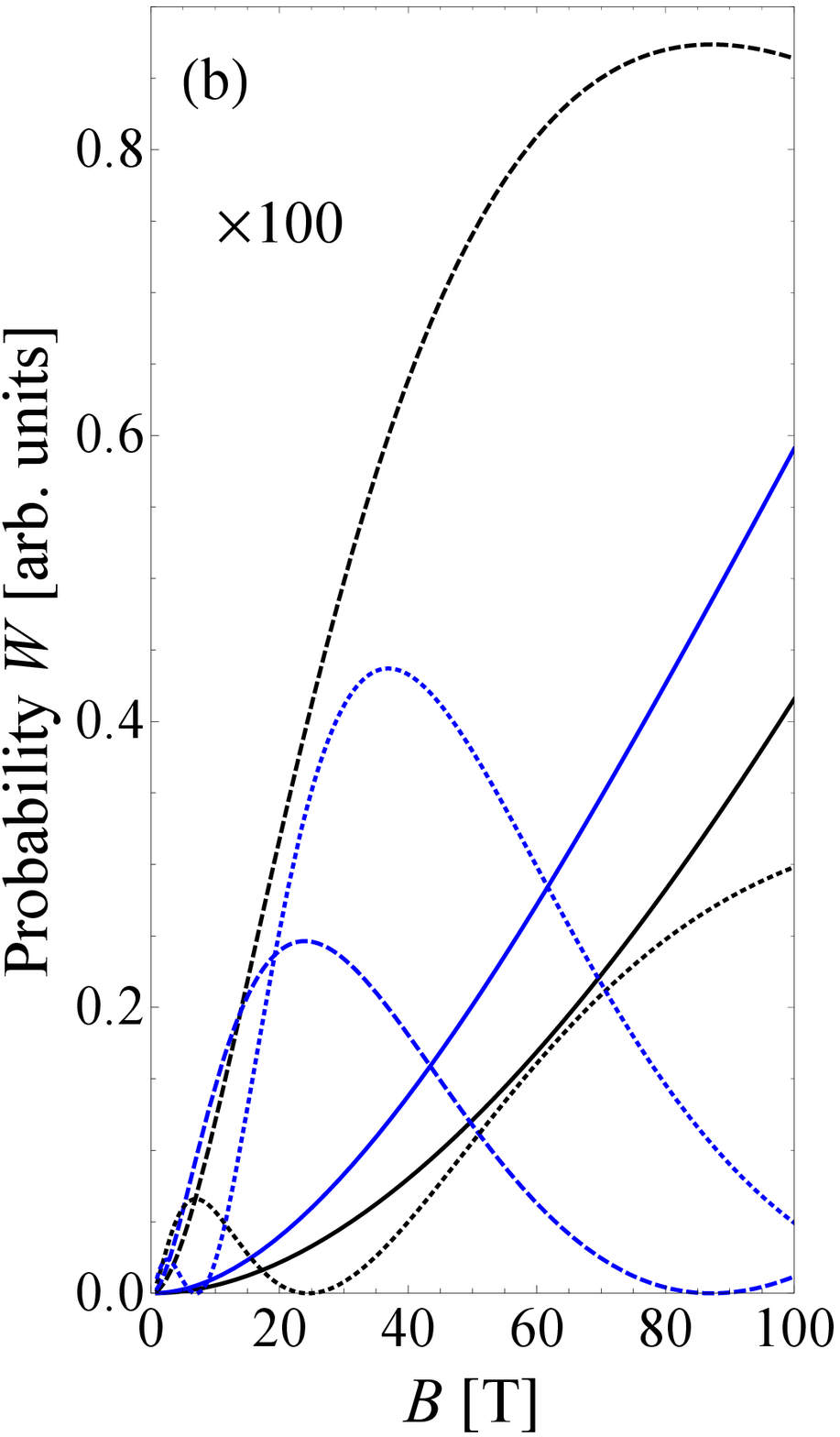}
\caption{Intensities of optical transitions from the valence band to impurity states as functions of pseudo-magnetic field strength $B$ [by Eq.~\eqref{EqFermiGoldRule}]. 
(a) Transitions from the valence band states with $l=0$, $n_b=0$ (black) and $n_b=1$ (blue) to the donor states with the angular quantum quantum number $m=1/2$ and the radial quantum numbers $n_i=0$ (solid), $n_i=1$ (dashed), and $n_i=2$ (dotted). 
(b) Transitions from the valence band states with $l=-1$, $n_b=0$ (black) and $n_b=1$ (blue) to the acceptor states with the angular momentum quantum number $m=-1/2$  and the radial quantum numbers $n_i=0$ (solid), $n_i=1$ (dashed), and $n_i=2$ (dotted).}
\label{Fig5}
\end{figure}
We observe a set of resonances in the frequency domain
$0.4 \,{\rm eV}<\omega$ in panels $(b,d,f)$.
They correspond to the transitions from the Landau levels of the valence band to the ground state of the acceptor, whereas the domain 
$1.1\,{\rm eV}<\omega$ in panels $(a,c,e)$ depicts the transitions to the donor ground state.
The height of each peak is proportional to the probability of corresponding band-impurity transition. 
We conclude that, first, the probabilities of the transitions to donor states are much larger than the ones to the acceptor states. 
Second, the probabilities of impurity-band transitions are very sensitive to the artificial magnetic field strength $B$.

Figure~\ref{Fig5} clearly demonstrates this dependence for transitions from several highest (in electron representation) quasi-Landau levels in valence band to several lowest states of donors and acceptors. 
Such transitions play an important role at low temperatures. 
Their intensities can be driven by the deformation of the TMD layer even at relatively small values of the pseudo-magnetic field (or the deformation parameter $b$).   
This allows us to use strain as an auxiliary degree of freedom utilized to monitor and control the optical transitions in 2D materials.

Let us now discuss the limitations of our analytical theoretical model of impurity-band transitions.
First, from the general perspective, we study a two-band model based on the Hamiltonians~\eqref{BareHamiltonian} and~\eqref{coulomb}, which conserve the azimuthal symmetry. 
That is why the impurity eigenstates are characterized by the angular momentum even in the presence of artificial uniform magnetic field. 
The possible presence of the warping terms~\cite{RefKormanyos} in the Hamiltonian would result in a mixing of the states with different angular momenta and thus, the impurity states, strictly speaking, represent linear superpositions of the states with given momenta. Second, we do not consider the intervalley mixing of impurity states.
However, because the warping terms are usually  small~\cite{RefKormanyos}, and the intervalley mixing is not sufficient due to the large distance between non-equivalent valleys in the reciprocal space, we neglected their influence on the optical transitions.
%
Third, we have disregarded the possible excitonic effects, the account of which requires a many-body treatment of the problem based on the Bethe-Salpeter equation~\cite{PhysRevLett.121.167402, RefGolubIvch}. 
A careful analysis of excitonic and warping-dependent effects requires non-analytic methods, especially, in the presence of the strain, that is beyond the scope of the present paper.


\section*{Conclusions} 
We studied the magneto-optic effect on impurity-band transitions in TMD monolayers under the action of artificial pseudo-magnetic field produced by an elastic deformation of the crystal.
The system demonstrates high sensitivity of both the resonant frequencies and the intensities of the corresponding impurity-band transition peaks in the spectrum to the applied stress (pseudomagnetic field). 
Taking concrete, typical for TMDs low-lying impurity states, we revealed the possibility to manipulate the optical properties of the system by the layer deformation, which opens opportunities for using strain-dependent pseudomagnetic fields as an experimental tool. 

\acknowledgements
The numerical part of the calculations in this paper was financially supported by the Russian Science Foundation (Project No.~17-12-01039); the analytical part of the work was supported by the Foundation for the Advancement of Theoretical Physics and Mathematics ``BASIS'', and the Institute for Basic Science in Korea (Project No.~IBSR024-D1). 




\bibliography{references}
\bibliographystyle{apsrev4-1}


\end{document}